\documentclass[iop]{emulateapj}

\pdfoutput=1
\usepackage[fleqn]{amsmath}
\usepackage{amsfonts}
\usepackage{amstext}
\usepackage{amssymb}
\usepackage[colorlinks=true, citecolor=blue, linkcolor=blue, breaklinks=true]{hyperref}
\usepackage{natbib}
\newcommand{\hi } {{\rm H}\,{\footnotesize\rm I} \,}
\newcommand{\hiA} {{\rm H}\,{\footnotesize\rm I}}
\newcommand{\hii } {{\rm H}\,{\footnotesize\rm II} \,}

\begin{document}
\nocite{*}

\title{SPARC: Mass models for 175 disk galaxies with\\ Spitzer Photometry And Accurate Rotation Curves.}
\author{Federico Lelli$^{1, *}$, Stacy S. McGaugh$^{1}$ and James M. Schombert$^{2}$}
\affil{$^{1}$Department of Astronomy, Case Western Reserve University, Cleveland, OH 44106, USA\\
$^{2}$Department of Physics, University of Oregon, Eugene, OR 97403, USA}
\email{$^{*}$federico.lelli@case.edu}

\begin{abstract}
We introduce SPARC (Spitzer Photometry \& Accurate Rotation Curves): a sample of 175 nearby galaxies with new surface photometry at 3.6~$\mu$m and high-quality rotation curves from previous \hiA/H$\alpha$ studies. SPARC spans a broad range of morphologies (S0 to Irr), luminosities ($\sim$5 dex), and surface brightnesses ($\sim$4 dex). We derive [3.6] surface photometry and study structural relations of stellar and gas disks. We find that both the stellar mass$-$\hi mass relation and the stellar radius$-$\hi radius relation have significant intrinsic scatter, while the \hi mass$-$radius relation is extremely tight.\\ 
We build detailed mass models and quantify the ratio of baryonic-to-observed velocity ($V_{\rm bar}/V_{\rm obs}$) for different characteristic radii and values of the stellar mass-to-light ratio ($\Upsilon_{\star}$) at [3.6]. Assuming $\Upsilon_{\star}\simeq 0.5$ $M_{\odot}/L_{\odot}$ (as suggested by stellar population models) we find that (i)~the gas fraction linearly correlates with total luminosity, (ii) the transition from star-dominated to gas-dominated galaxies roughly corresponds to the transition from spiral galaxies to dwarf irregulars in line with density wave theory; and (iii)~$V_{\rm bar}/V_{\rm obs}$ varies with luminosity and surface brightness: high-mass, high-surface-brightness galaxies are nearly maximal, while low-mass, low-surface-brightness galaxies are submaximal. These basic properties are lost for low values of $\Upsilon_{\star} \simeq 0.2$ $M_{\odot}/L_{\odot}$ as suggested by the DiskMass survey. The mean maximum-disk limit in bright galaxies is $\Upsilon_{\star}\simeq 0.7$ $M_{\odot}/L_{\odot}$ at [3.6]. The SPARC data are publicly available and represent an ideal test-bed for models of galaxy formation.
\end{abstract}


\section{Introduction}\label{sec:intro}

The study of rotation curves has played a key role in our understanding of the mass distribution in galaxies \citep[e.g.,][]{Rubin1978, Bosma1981, vanAlbada1985}. Atomic hydrogen (\hiA) is one of the best kinematical tracer of the gravitational potential in nearby galaxies for two basic reasons:
\begin{enumerate}
 \item \hi gas is dynamically cold and follows nearly circular orbits, hence it directly traces the gravitational potential. The \hi velocity dispersion is typically of $\sim$10 km s$^{-1}$, so pressure support becomes significant only in very low-mass galaxies with rotation velocities of $\sim$20 km~s$^{-1}$ \citep[e.g.,][]{Lelli2012b}. Strong assumptions about the velocity tensor, which are necessary in studies of stellar kinematics, are not required in \hi studies.
 \item \hi gas is diffuse and extends well beyond the galaxy stellar component, hence the gravitational potential can be traced up to several effective radii, where dark matter (DM) is expected to dominate.
\end{enumerate}
Studies of \hi rotation curves, however, have been historically limited to small galaxy samples because they are significantly time costly: they require interferometric \hi observations and careful modelling of 3D datasets.

\begin{table*}
\begin{center}
\setlength{\tabcolsep}{3.5pt}
\caption{Galaxy Sample}
\begin{tabular}{lcccccccccccccc}
\hline
\hline
Name & $T$ & $D$ & Met. & $i$ &$L_{[3.6]}$ & $R_{\rm eff}$ & $\Sigma_{\rm eff}$ & $R_{\rm d}$ & $\Sigma_{d}$ & $M_{\hi}$ & $R_{\hi}$ & $V_{\rm{f}}$ & $Q$ & Ref. \\
     & & (Mpc)&        & ($^{\circ}$) & ($10^{9} L_{\odot}$) & (kpc) & ($L_{\odot}$ pc$^{-2}$) & (kpc) & ($L_{\odot}$ pc$^{-2}$) & ($10^{9} M_{\odot}$) & (kpc) & (km s$^{-1}$) & & \\
(1)   &(2) & (3)           &(4)& (5)      & (6)             & (7)  & (8)   & (9)  & (10)   & (11)   &(12) & (13) &(14) & (15)\\
\hline
CamB   &10 & 3.36$\pm$0.26 & 2 & 65$\pm$5 & 0.075$\pm$0.003 & 1.21 &  7.9 & 0.47 &  66.2 &  0.016 & ... &  ... & 2 & 1\\
D512-2 &10 & 15.20$\pm$4.56& 1 & 56$\pm$10& 0.325$\pm$0.022 & 2.37 &  9.2 & 1.24 &  93.9 &  0.108 & ... & ... & 3 & 2\\
D564-8 &10 & 8.79$\pm$0.28 & 2 & 63$\pm$7 & 0.033$\pm$0.004 & 0.72 & 10.1 & 0.45 &  30.8 &  0.039 & ... & ... & 3 & 2\\
D631-7 &10 & 7.72$\pm$0.18 & 2 & 59$\pm$3 & 0.196$\pm$0.009 & 1.22 & 20.9 & 0.70 & 115.0 &  0.385 & ... & 57.7$\pm$2.7 & 1 & 2, 3\\
\hline
\end{tabular}
\tablecomments{Table \ref{tab:sample} is published in its entirety in a machine readable format. A portion is shown here for guidance regarding its form and content. References for \hi and H$\alpha$ data: (1) \citet{Begum2003}; (2) \citet{Trachternach2009}; (3) \citet{deBlok2001}.}
\label{tab:sample}
\end{center}
\end{table*}
The mass modelling of galaxies is generally plagued by the ``disk-halo degeneracy'' \citep{vanAlbada1985}: the relative contributions of stellar disk and DM halo to the observed rotation curve are strongly degenerate due to uncertainties in the stellar mass-to-light ratio ($\Upsilon_{\star}$). More generally, this should be called ``star-halo degeneracy'' because it also applies to bulge-dominated spirals \citep[e.g.,][]{Noordermeer2006} and elliptical galaxies \citep[e.g.,][]{Oguri2014}. To minimize the star-halo degeneracy, the best approach is to use near-infrared (NIR) surface photometry ($K$-band or 3.6 $\mu$m), which provides the closest proxy to the stellar mass \citep[e.g.,][]{Verheijen2001b}. Stellar populations synthesis (SPS) models suggest that $\Upsilon_{\star}$ displays much smaller variations in the NIR than in optical bands and depends weakly on the star formation history of the galaxy. Several models actually predict that $\Upsilon_{\star}$ is nearly constant in the NIR over a broad range of galaxy masses and morphologies \citep[e.g.,][]{Bell2001, Portinari2004, Meidt2014, Schombert2014a, McGaugh2014}. The overall normalization of $\Upsilon_{\star}$, however, is still uncertain up to a factor of $\sim$3 due to differences in the SPS models (such as the treatment of asymptotic-giant-branch stars) and in the assumed stellar initial mass function (IMF). These issues are described in detail in Sect.~\ref{sec:ML}.

To advance our understanding of the mass distribution in galaxies, we have built SPARC (Spitzer Photometry and Accurate Rotation Curves): a sample of 175 galaxies with new surface photometry at 3.6 $\mu$m and extended \hi rotation curves from the literature. For $\sim$1/3 of SPARC galaxies, we also have H$\alpha$ rotation curves from long-slit spectroscopy, integral field units (IFUs), and Fabry-Perot interferometry, probing the inner regions at high spatial resolutions. Contrary to previous samples, SPARC spans very broad ranges in luminosity, surface brightness, rotation velocity, and Hubble type, forming a representative sample of disk galaxies in the nearby Universe. To date, this is the largest collection of galaxies with both high-quality rotation curves and NIR surface photometry.

The SPARC data will be explored in a series of papers and are available on-line at \url{astroweb.cwru.edu/SPARC}. Preliminary SPARC data have already been used in several publications: (i) \citet{Lelli2016} study the baryonic Tully-Fisher relation (BTFR); (ii) \citet{DiCintio2016} investigate the mass discrepancy-acceleration relation in a $\Lambda$CDM context; and (iii) \citet{Katz2016} fit different DM halo profiles to the SPARC rotation curves. In this paper, we present the galaxy sample (Sect.~\ref{sec:sample}), derive surface photometry at 3.6 $\mu$m (Sect.~\ref{sec:photo}), describe the collection of \hiA/H$\alpha$ rotation curves (Sect.~\ref{sec:rotcur}), build mass models (Sect.\ref{sec:massmodels}), study stellar and \hi scaling relations (Sect.~\ref{sec:scaling}), and finally investigate gas fractions (Sect.~\ref{sec:gasfrac}) and the degree of baryonic maximality (Sect.~\ref{sec:maximality}) for different values of $\Upsilon_{\star}$.

\section{Galaxy Sample}\label{sec:sample}

We collected more than 200 extended \hi rotation curves from previous compilations, large surveys, and individual studies (see Sect.~\ref{sec:rotcur} for details). This kinematic dataset is the result of $\sim$30 years of interferometric \hi observations using the Westerbork Synthesis Radio Telescope (WSRT), Very Large Array (VLA), Australia Telescope Compact Array (ATCA), and Giant Metrewave Radio Telescope (GMRT). Subsequently, we searched the Spitzer archive and found useful [3.6] images for 175 galaxies. Most of these objects are part of the Spitzer Survey for Stellar Structure in Galaxies \citep[S$^{4}$G,][]{Sheth2010}. We also used [3.6] images from \citet{Schombert2014b} for low surface brightness (LSB) galaxies, which are usually under-represented in optical catalogues with respect to high surface brightness (HSB) spirals \citep[e.g.,][]{McGaugh1995b}. SPARC is not a statistically complete or volume-limited sample, but it forms a representative sample of disk galaxies in the nearby Universe spanning the widest possible range of properties for galaxies with extended rotation curves.

\begin{figure*}[thb]
\centering
\includegraphics[width=\textwidth]{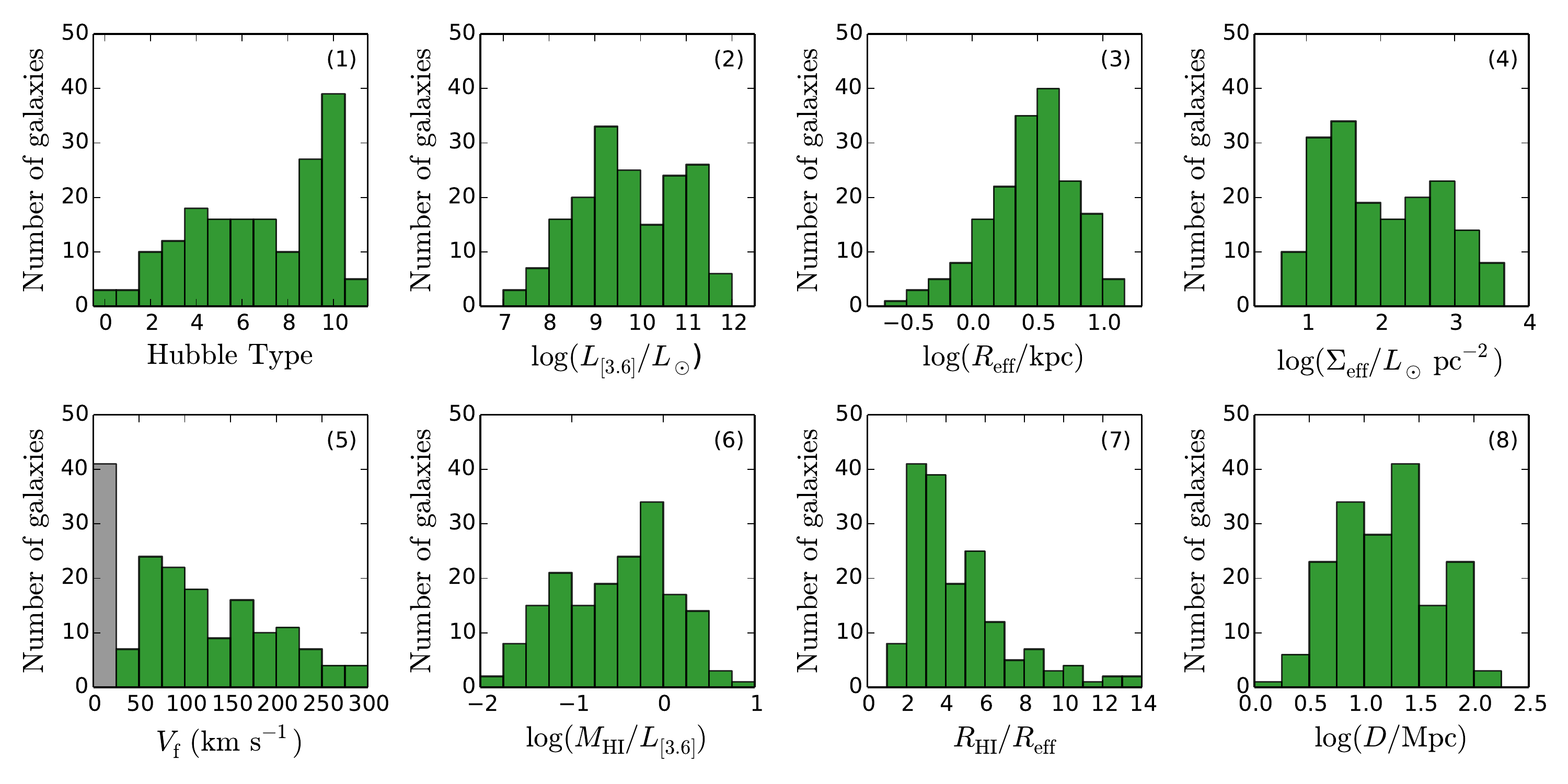}
\caption{Overall properties of SPARC galaxies: (1) numerical Hubble type; (2) total [3.6] luminosity; (3) effective [3.6] surface brightness; (4) effective radius; (5) velocity along the flat part of the rotation curve; we assign $V_{\rm f}=0$ to rotation curves that do not reach a flat part within $\sim$5$\%$ (grey bin); (6) ratio of \hi mass to [3.6] luminosity; (7) ratio of \hi radius to effective radius; (8) assumed distance.}
\label{fig:histo}
\end{figure*}

Table~\ref{tab:sample} lists the main properties of SPARC galaxies.\\
\textit{Column} (1) gives the galaxy name.\\
\textit{Column} (2) gives the numerical Hubble type from \citet{RC3}, \citet{Schombert1992}, or NED\footnote{The NASA/IPAC Extragalactic Database (NED) is operated by the Jet Propulsion Laboratory, California Institute of Technology, under contract with the National Aeronautics and Space Administration.} adopting the following scheme: 0~=~S0, 1~=~Sa, 2~=~Sab, 3~=~Sb, 4~=~Sbc, 5~=~Sc, 6~=~Scd, 7~=~Sd, 8~=~Sdm, 9~=~Sm, 10~=~Im, 11~=~BCD. \\
\textit{Column} (3) gives the assumed distance (see Sect.~\ref{sec:dist}).\\
\textit{Column} (4) gives the distance method: 1~=~Hubble flow, 2~=~Tip of the red giant branch, 3~=~Cepheids, 4~=~Ursa Major cluster of galaxies, 5~=~Supernovae.\\
\textit{Column} (5) gives the assumed inclination angle ($i$).\\
\textit{Column} (6) gives the total luminosity at 3.6 $\mu$m ($L_{[3.6]}$).\\
\textit{Column} (7) gives the effective radius ($R_{\rm eff}$) encompassing half of the total luminosity.\\
\textit{Column} (8) gives the effective surface brightness ($\Sigma_{\rm eff}$).\\
\textit{Column} (9) gives the scale length of the stellar disk ($R_{\rm d}$).\\
\textit{Column} (10) gives the extrapolated central surface brightness of the stellar disk ($\Sigma_{\rm d}$).\\
\textit{Column} (11) gives the total \hi mass ($M_{\hi}$).\\
\textit{Column} (12) gives the \hi radius ($R_{\hi}$) where the \hi surface density (corrected to face-on) reaches 1 $M_{\odot}$~pc$^{-2}$.\\
\textit{Column} (13) gives the velocity along the flat part of the rotation curve ($V_{\rm f}$) derived as in \citet{Lelli2016}.\\
\textit{Column} (14) gives the rotation-curve quality flag ($Q$): 1~=~high, 2~=~medium, 3~=~low. See Sect.~\ref{sec:rotcur} for details.\\
\textit{Column} (15) gives the reference for the \hi data.

Inclination angles are typically derived by fitting a tilted-ring model to the \hi velocity fields \citep{Begeman1987} and considering systematic variations with radius (warps): column 5 provides the mean value of $i$ in the outer parts of the \hi disk. In some cases, \hi velocity fields are not adequate to properly trace the run of $i$ with radius, hence the inclination is fixed to the optical value \citep[e.g.,][]{deBlok1996}. The derivation of photometric properties (columns 6 to 10) is described in Sect.~\ref{sec:photo}.

\subsection{General Properties}\label{sec:prop}

The overall properties of SPARC galaxies are shown with histograms in Figure~\ref{fig:histo}. SPARC spans a broad range in morphologies (S0 to Im/BCD), luminosities ($\sim$10$^7$ to $\sim$10$^{12}$ $L_{\odot}$), effective radii ($\sim$0.3 to $\sim$15 kpc), effective surface brightnesses ($\sim$5 to $\sim$5000 $L_{\odot}$~pc$^{-2}$), rotation velocities ($\sim$20 to $\sim$300 km~s$^{-2}$), and gas content ($0.01 \lesssim M_{\hi}/L_{[3.6]}\lesssim 10$). Galaxies with $10^{10} < L_{[3.6]}/L_{\odot} < 10^{10.5}$ are slightly under-represented (in a volume-weighted sense): apparently \hi studies have tended to focus on bright spirals and dwarf galaxies. As expected, SPARC does not contain galaxies with $L_{[3.6]} > 10^{12}$~$L_{\odot}$ which typically are gas-poor ellipticals in clusters, nor galaxies with $L_{[3.6]} < 10^{7}$~$L_{\odot}$ for which our knowledge is restricted to gas-poor spheroidals around the Milky Way and M31.

Most galaxies in SPARC have late Hubble types (9 to 10), corresponding to Sm and Im galaxies. This happens because the Hubble classification is intrinsically biased towards HSB features and assign most LSB galaxies within these two latest stages \citep{McGaugh1995a}. Currently, the SPARC dataset contains only three lenticulars because early-type galaxies generally lack a high-density \hi disk and have been historically neglected by interferometric \hi studies. Recent surveys, however, show that several early-type galaxies in the group and field environments possess an outer low-density \hi disk \citep{Serra2012}, which can be used for kinematic investigations \citep{denHeijer2015}.

The ratio $R_{\hiA}/R_{\rm eff}$ varies from $\sim$1 to $\sim$14 (Panel 8 of Fig.~\ref{fig:histo}). Clearly, \hi rotation curve probes the gravitational potential out to much larger radii than existing IFU surveys like CALIFA \citep{GarciaLorenzo2015} or MANGA \citep{Bundy2015}. Historically, the relative extent of \hiA-to-stellar disks has been measured using either $R_{\hiA}/R_{25}$ where $R_{25}$ is the B-band isophotal radius at 25 mag~arcsec$^{-2}$ \citep{Broeils1997, Noordermeer2005}, or $R_{\hiA}/R_{\star}$ where $R_{\star} = 3.2 R_{\rm d}$ \citep{Swaters2009, Lelli2014}. The latter definition allows comparing stellar disks with different central surface brightnesses: a pure exponential disk with $\Sigma_{\rm d} = 21.5$ B mag arcsec$^{-2}$ \citep{Freeman1970} has $R_{\star} = R_{25}$. We find $\langle R_{\hiA}/R_{\star} \rangle= 1.9 \pm 1.0$ with a median value of 1.7. This is in line with previous estimates of $\langle R_{\hiA}/R_{\star} \rangle$ for different galaxy types:  $1.7 \pm 0.7$ \citep{Noordermeer2005} for early-type galaxies (S0 to Sab), $1.7 \pm 0.5$ \citep{Broeils1997} for spiral galaxies (mostly Sb to Scd), $1.8 \pm 0.8$ \citep{Swaters2009} for late-type dwarfs (Sd to Im), and $1.7 \pm 0.5$ \citep{Lelli2014} for starburst dwarfs (Im and BCD). \citet{Martinsson2016}, however, find $R_{\hi}/R_{25}=1.35 \pm 0.22$ for 28 late-type spirals. We note that SPARC galaxies are representative of the disk population in the field, nearby groups, and diffuse clusters like Ursa Major \citep{Tully1997}. Galaxies in dense clusters like Virgo may show smaller values of $R_{\hiA}/R_{\star}$ due to gas stripping \citep[e.g.,][]{Chung2009}.

\subsection{Galaxy Distances}\label{sec:dist}

The galaxies in SPARC can be classified in three main groups depending on the distance estimate.

\textit{I. Accurate Distances}: 50 objects have distances from the tip of the red giant branch (45), Cepheids (3), or supernovae (2). These distances have errors ranging from $\sim$5$\%$ to $\sim$10$\%$ and are generally on the same zero-point scale \citep[see discussion in][]{Tully2013}. The vast majority of these distances are drawn from the Extragalactic Distance Database \citep{Jacobs2009, Tully2013}.

\textit{II. Ursa Major}: 28 objects lie in the Ursa Major cluster \citep{Verheijen2001a}, which has an average distance of $18 \pm 0.9$ Mpc \citep{Sorce2013}. This explains the peak at $\log(D/\mathrm{Mpc})\simeq1.25$ in panel 10 of Fig.~\ref{fig:histo}. The Ursa Major cluster has an estimated depth of $\sim$2.3~Mpc \citep{Verheijen2001b}, hence we adopt an error of $\sqrt{(2.3^2 + 0.9^2)} = 2.5$ Mpc for individual galaxies. Two galaxies from \citet{Verheijen2001b} are considered as background/foreground objects: NGC~3992 having $D\simeq24$ Mpc from a Type-Ia supernova \citep{Parodi2000}, and UGC~6446 laying near the cluster boundary in both space and velocity ($D\simeq12$ Mpc from the Hubble flow).

\textit{III. Hubble Flow}: 97 objects have Hubble-flow distances corrected for Virgocentric infall (taken from NED). We assume $H_{0} = 73$ km~s$^{-1}$~Mpc$^{-1}$, giving distances on a similar zero-point scale as those in groups I and II \citep[e.g.,][]{Tully2013}. Hubble-flow distances are very uncertain for nearby galaxies where peculiar velocities may strongly contribute to the systemic velocities, but become more accurate for distant objects. We adopt the following error scheme: 30$\%$ for $D \leq 20$~Mpc; 25$\%$ for $20 < D \leq 40$~Mpc; 20$\%$ for $40 < D \leq 60$~Mpc; 15$\%$ for $60 < D \leq 80$~Mpc; and 10$\%$ for $D > 80$~Mpc. This scheme considers that peculiar velocities can be as high as $\sim$500~km~s$^{-1}$ and $H_{0}$ has an uncertainty of $\sim$7$\%$. \citet{Lelli2016} show that the intrinsic scatter in the BTFR for galaxies in groups I+II is similar to that for the entire sample, indicating that the assumed errors on Hubble-flow distances are realistic.

\section{Data Collection \& Analysis}

\subsection{Surface Photometry}\label{sec:photo}

For the 175 galaxies in SPARC, we derived homogeneous surface photometry at 3.6~$\mu$m following the procedures of \citet{Schombert2014b}. The frames were flat-fielded and calibrated using the standard Spitzer pipeline. Subsequently, they were interactively cleaned by masking bright contaminating sources, like foreground stars and background galaxies. This is a crucial step because [3.6] images contain nearly 10 times more point sources than similar exposures in optical bands. Masked pixels were subsequently replaced with the local mean isophotal value for aperture luminosity determination.

The cleaned frames were analysed using the Archangel software \citep{Schombert2011}. The absolute sky value and its error were estimated using multiple sky-boxes, i.e. regions in the frame that are free of background/foreground objects. The standard Spitzer pipeline provides a formal zero-point error smaller than 2$\%$, but a more realistic photometric error should also consider the sky brightness error, which dominates all other sources of uncertainty in the surface photometry and aperture magnitudes \citep{Schombert2014b}. The mean error on sky brightness is $\sim$3$\%$.

To derive azimuthally averaged surface brighntess profiles, we fitted ellipses to the frames using a standard Fourier-series least-square algorithm. During the ellipse fitting, pixels with values above 3$\sigma$ of the mean isophote were automatically masked and replaced with the mean value at that radius. These pixels typically corresponds to unresolved background/foreground objects, but we visually inspected each frame to ensure that real asymmetric features in the galaxy (like prominent spiral arms or \hii regions) were not removed. Most low-mass and LSB galaxies have irregular morphologies, hence ellipses only provide a zeroth-order apporximation of their shape. The errors in the surface brightness profiles, however, are dominated by the uncertainties in the sky value rather than variations around each ellipse. For thin, egde-on galaxies, ellipse fitting does not provide satisfactory results in the central regions, hence we derived surface brightness profiles considering a narrow circular section around the semi-major axes and average both sides.

To estimate aperture magnitudes, we built curves-of-growth by summing both raw and replaced pixels within the best-fit ellipse at a given radius (sub-pixels are used at the ellipse edges). The amount of light from replaced pixels varies from 2$\%$ to 15$\%$. Summing pixels generally leads to unstable curves-of-growth at large radii, where the galaxy surface brightness is comparable to the contaminating halos of imperfectly masked sources. Hence, we replaced the aperture values with the mean isophotal values at large radii (beyond the radius containing $\sim$90$\%$ of the raw luminosity). This provides curves-of-growth that approach a clear asymptotic value at large radii. We calculated total magnitudes using asymptotic fits to the curves-of-growth. Corresponding errors were estimated by recalculating the total magnitudes using a 1$\sigma$ variation in the sky value. Total luminosities assume a solar absolute magnitude of 3.24 at 3.6 $\mu$m \citep{Oh2008}. We also calculated the effective radius ($R_{\rm eff}$), i.e. the radius encompassing half of the total luminosity, and the effective surface brightness ($\Sigma_{\rm eff}$), i.e. the average surface brightness within $R_{\rm eff}$.

To estimate the disk scale length ($R_{\rm d}$) and central surface brightness ($\mu_{\rm d}$), we fitted exponential functions to the outer parts of the surface brightness profiles. We stress that the results depend on the fitted radial range, which may be ambiguous for galaxies with outer down-bending (type II) or up-bending (type III) profiles \citep[cf.][]{Erwin2008}. For most low-mass and LSB galaxies, exponential functions provide a good description of the surface brightness profiles over a broad radial range but several galaxies show light enanchements/depressions in the inner parts \citep[similar to optical bands, e.g.,][]{Swaters2002b, Schombert2011b}. As expected, the vast majority of spiral galaxies show deviations from exponential profiles in the inner parts due to bulges, bars, and lenses. All surface brightness profiles are available for download at \url{astroweb.cwru.edu/SPARC}.

\subsection{Rotation Curves}\label{sec:rotcur}

The majority ($\sim$75$\%$) of SPARC rotation curves are the result of Ph.D. theses from the University of Groningen \citep{Begeman1987, Broeils1992, deBlok1997, Verheijen1997, Swaters1999, Noordermeer2006, Lelli2013}. They were derived using similar techniques and software, forming a relatively homogeneous dataset. These 175 rotation curves are drawn from the following publications (in parenthesis we provide the corresponding number of objects excluding duplicates): \citet[][32]{Swaters2009}, \citet[][30]{Sanders1998}, \citet[][17]{Sanders1996}, \citet[][12]{Noordermeer2007}, \citet[][10]{deBlok2001}, \citet[][9]{Begeman1991}, \citet[][8]{Lelli2012a, Lelli2012b, Lelli2014}, \citet[][7]{deBlok2002}, \citet[][7]{deBlok1996}, \citet[][6]{Kuzio2008}, \citet[][5]{Spekkens2006}, \citet[][4]{Gentile2004, Gentile2007}, \citet[][4]{vanZee1997}, \citet[][3]{Cote2000}, \citet[][3]{Fraternali2002, Fraternali2011}, \citet[][2]{Hallenbeck2014}, \citet[][2]{Trachternach2009}, \citet[][1]{Barbieri2005}, \citet[][1]{Battaglia2006}, \citet[][1]{Begum2003}, \citet[][1]{Begum2004}, \citet[][1]{Blais2004}, \citet[][1]{Boomsma2008}, \citet[][1]{Chemin2006}, \citet[][1]{Elson2010}, \citet[][1]{Kepley2007}, \citet[][1]{Richards2015}, \citet[][1]{Simon2003}, \citet[][1]{VerdesMontenegro1997}, \citet[][1]{Verheijen1999}, and \citet[][1]{Walsh1997}. For duplicate galaxies, we select the most reliable rotation curve. Table~\ref{tab:sample} provides the reference for each galaxy.

We do not consider rotation curves from THINGS \citep{deBlok2008} and LITTLE-THINGS \citep{Oh2015} for the sake of homogeneity. These rotation curves are characterized by many small-scale bumps and wiggles that are not observed in other rotation curve samples. Most likely these small-scale features do not trace the smooth, axisymmetric gravitational potential, but may be due to non-circular components like streaming motions along spiral arms \citep[e.g.,][]{Khoperskov2015}. SPARC rotation curves are generally smooth but can show large-scale features with a direct correspondence in the surface brightness profile, in agreement with Renzo's rule \citep{Sancisi2004}: ``For any feature in the luminosity profile there is a corresponding feature in the rotation curve and vice versa''.

\subsubsection{Beam smearing effects and hybrid rotation curves}

The inner parts of rotation curves are a key tool to distinguish between cuspy or cored DM profiles \citep[e.g.,][]{deBlok2001, deBlok2008, Gentile2004}, but they must be derived with great care because the limited spatial resolution may lead to a systematic underestimate of the inner rotation velocities \citep[the so-called beam-smearing effects, see e.g.,][]{Swaters2009}. Several techniques have been developed to take beam-smearing effects into account: (i) projecting the derived rotation curves onto position-velocity diagrams and correcting the inner velocity points by visual inspection \citep[e.g.,][]{Begeman1987, deBlok1996, Verheijen2001a}, and (ii) building 3D disk models and comparing model-cubes with observed cubes \citep[e.g.,][]{Swaters2009, Lelli2010}. We carefully checked each rotation curve and conclude that beam smearing does not play a major role. For only a few galaxies, the innermost points were perhaps affected by beam smearing and have been rejected.

For 56 galaxies ($\sim$1/3 of SPARC), we have hybrid rotation curves combining H$\alpha$ data in the inner regions with \hi data in the outer parts. H$\alpha$ rotation curves trace the kinematics at high spatial resolutions ($\sim$1$''$), hence beam-smearing effects are minimal. Hybrid rotation curves were derived using H$\alpha$ IFU observations \citep[7 objects from][]{Kuzio2008, Richards2015} and long-slit spectroscopy \citep[41 objects from][]{Roelfsema1985, Walsh1997, Cote2000, deBlok2001, deBlok2002, Gentile2004, Spekkens2006, Noordermeer2007}. The raw H$\alpha$ data are typically fit with a spline to derive smooth rotation curves \citep{deBlok2001, deBlok2002, Noordermeer2007}, which average over small-scale irregularities and trace the overall gravitational potential. We built another 8 hybrid rotation curves using H$\alpha$ data from Fabry-Perot interferometry \citep{Blais1999, Blais2001, Blais2004, Daigle2006, Dicaire2008}. Apart a few cases (notably NGC\,24), \hi and H$\alpha$ rotation curves are generally in good agreement, indicating that the former were properly determined \citep[see also][]{McGaugh2001, Swaters2009}. The rotation curve of NGC~2976 combines H$\alpha$ and CO data \citep{Simon2003}.

Despite the high spatial resolution, H$\alpha$ rotation curves are sometimes quite uncertain due to the patchy distribution and complex kinematics of H$\alpha$ gas, especially for low-mass and LSB galaxies. We use only \hi data for the several objects: DDO~154, IC~2574, NGC~3109, NGC~5033, NGC~5055, NGC~6946, NGC~7331, UGC~2259. \hi and H$\alpha$ rotation curves, however, agree within the errors.

\begin{table}
\begin{center}
\caption{Mass Model for UGCA~442}
\begin{tabular}{ccccccc}
\hline
\hline
Rad   & $V_{\rm obs}$ & $V_{\rm gas}$ & $V_{\rm disk}$ & $V_{\rm bul}$ & $\Sigma_{\rm disk}$ & $\Sigma_{\rm bul}$ \\
(kpc) & \multicolumn{4}{c}{------------ (km~s$^{-1}$) ------------}    & \multicolumn{2}{c}{-- ($L_{\odot}$ pc$^{-2}$) --}\\
\hline
0.42 & 14.2$\pm$1.9 & 4.9  & 4.8  & 0.0 & 11.0 & 0.0 \\
1.26 & 28.6$\pm$1.8 & 13.1 & 10.8 & 0.0 & 5.8  & 0.0 \\
2.11 & 41.0$\pm$1.7 & 19.6 & 13.6 & 0.0 & 2.7  & 0.0 \\
2.96 & 49.0$\pm$1.9 & 22.4 & 13.3 & 0.0 & 1.0  & 0.0 \\
3.79 & 54.8$\pm$2.0 & 22.8 & 12.6 & 0.0 & 0.7  & 0.0 \\
4.65 & 56.4$\pm$3.1 & 21.4 & 12.3 & 0.0 & 0.4  & 0.0 \\
5.48 & 57.8$\pm$2.8 & 18.7 & 12.0 & 0.0 & 0.2  & 0.0 \\
6.33 & 56.5$\pm$0.6 & 16.7 & 10.6 & 0.0 & 0.0  & 0.0 \\
\hline
\end{tabular}
\tablecomments{$V_{\rm disk}$ and $V_{\rm bul}$ are given for $\Upsilon_{\star}=1$~M$_{\odot}$/L$_{\odot}$. Similar tables are available for all galaxies at \url{astroweb.cwru.edu/SPARC}.}
\label{tab:massmodel}
\end{center}
\end{table}
\subsubsection{Error budget and rotation curve quality}

The errors on the rotation velocities are estimated as in \citet{Swaters2009}:
\begin{equation}\label{eq:Verr}
 \delta_V^2 = \sqrt{\delta_{V_{\rm fit}}^2 + \bigg(\dfrac{V_{\rm app} - V_{\rm rec}}{4}\bigg)^{2}}
\end{equation}
where $V_{\rm app}$ and $V_{\rm rec}$ are the rotation velocities obtained from fitting separately the approaching and receding sides of the disk, respectively, and $\delta_{V_{\rm fit}}$ is the formal error from fitting the entire disk. We consider only velocity points that are derived from both sides of the disk. Eq.~\ref{eq:Verr} considers both local deviations from circular motions and global kinematic asymmetries between the two sides of the disk. It does not include, however, systematic errors due the assumed inclination. If the \hi disk is \textit{not} strongly warped, a change in $i$ would change the normalization of the rotation curve without altering its shape. Inclination corrections go as $\sin(i)$, thus they are small for edge-on galaxies but becomes large for face-on ones.

We assign a quality flag ($Q$) to each rotation curve using the following scheme: $Q=1$ for galaxies with high quality \hi data or hybrid H$\alpha$/\hi rotation curves (99 objects); $Q=2$ for galaxies with minor asymmetries and/or \hi data of lower quality (64 objects); $Q=3$ for galaxies with major asymmetries, strong non-circular motions, and/or off-sets between \hi and stellar distributions (12 objects). Galaxies with $Q=3$ are not suited for detailed dynamical studies: we build mass models for completeness but do not consider them in our analysis. Similarly, we provide mass models for face-on galaxies with $i<30^{\circ}$ but exclude them in our analysis. This does not introduce any selection bias since galaxy disks are randomly oriented on the sky. Our final science sample is made of 153 galaxies.

\begin{figure*}[thb]
\centering
\includegraphics[width=0.9\textwidth]{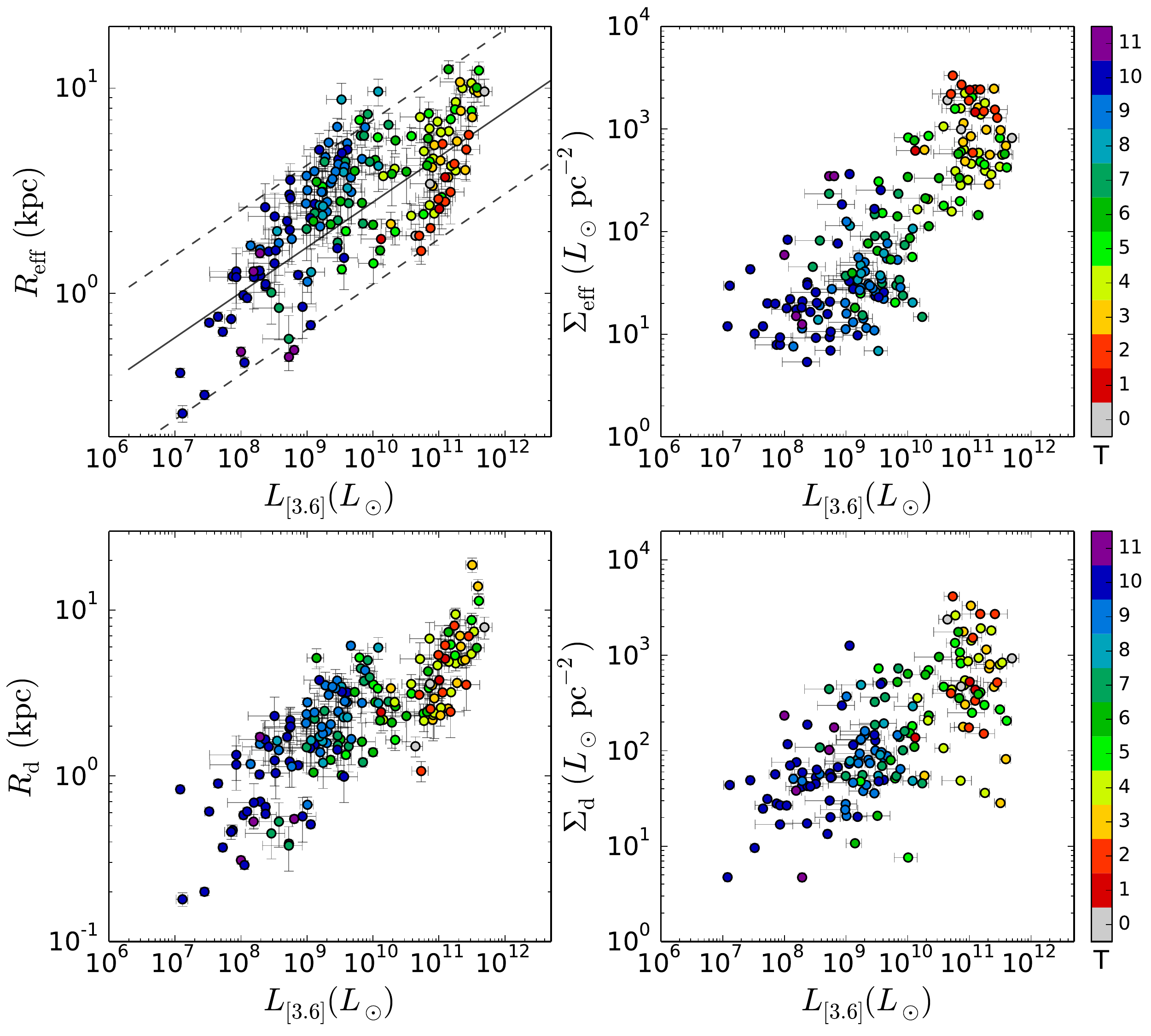}
\caption{Photometric relations. Galaxies are color-coded by Hubble type. In the top-left panel, the solid line shows the $K$-band mass-radius relation from the GAMA survey \citep{Lange2015}, where stellar masses have been converted to [3.6] luminosities assuming $\Upsilon_{\star} = 0.5$ $M_{\odot}/L_{\odot}$. The dashed lines show a 0.4 dex deviation from the mean relation, corresponding to $\sim$2$\sigma$.}
\label{fig:photo}
\end{figure*}
\subsection{Mass Models}\label{sec:massmodels}

We derive the contributions of gas disk ($V_{\rm gas}$), stellar disk ($V_{\rm disk}$), and bulge ($V_{\rm bul}$) to the observed rotation curves. The total baryonic contribution is then given by
\begin{equation}
 V_{\rm bar} = \sqrt{|V_{\rm gas}|V_{\rm gas} + \Upsilon_{\rm disk}|V_{\rm disk}|V_{\rm disk} + \Upsilon_{\rm bul}|V_{\rm bul}|V_{\rm bul}},
\end{equation}
where $\Upsilon_{\rm disk}$ and $\Upsilon_{\rm bul}$ are the stellar mass-to-light ratios of disk and bulge components, respectively. Note that absolute values are needed because $V_{\rm gas}$ can sometimes be negative in the innermost regions: this occurs when the gas distribution has a significant central depression and the material in the outer regions exerts a stronger gravitational force than that in the inner parts. In this paper, we investigate the ratio $V_{\rm bar}/V_{\rm obs}$ for different assumptions of $\Upsilon_{\star}$ (Sect.~\ref{sec:maximality}). Rotation curve fits with DM halos are presented in \citet{Katz2016} and will be exploited in future publications.

The gas contribution is calculated using the formula from \citet{Casertano1983}, which solves the Poisson equation for a disk with finite thickness and arbitrary density distribution. We assume a thin disk with a total mass of 1.33~$M_{\hi}$, where the factor 1.33 considers the contribution of helium. $V_{\rm gas}$ is either computed using \hi surface density profiles or taken from published mass models. For D512-2, D564-8, and D631-7, \hi surface density profiles are not available, hence we compute $V_{\rm gas}$ assuming an exponential profile with a scale length of 2~$R_{\rm d}$. This is a reasonable approximation for such late-type galaxies \citep[e.g.,][]{Swaters2002a}. We neglect the contribution of molecular gas because CO data are not available for most SPARC galaxies. Fortunately, molecules constitute a minor dynamical component and are generally sub-dominant with respect to stars and atomic gas \citep{Simon2003, Frank2016}. If molecules are distributed in a similar way as the stars, their contribution is implicitly included in the assumed value of $\Upsilon_{\star}$.

The stellar contribution is calculated using the observed [3.6] surface brightness profile and extrapolating the exponential fit to $R\rightarrow\infty$, unless the profile shows a clear truncation. For galaxies with $T<4$ (Hubble types earlier than Sbc), we perform bulge-disk decompositions in a non-parametric way. Firstly, we identify a fiducial radius $R_{\rm bul}$ within which the bulge light dominates. We then subtract the stellar disk from the profile by extrapolating an exponential fit at $R < R_{\rm bul}$: this fit describes the inner disk structure (like lenses, rings) and does not necessarily correspond to the outer disk fit. Finally, the residual luminosity profile is used to compute $V_{\rm bul}$ and linearly extrapolated at $R > R_{\rm bul}$ to avoid unphysical truncations. We assume that the bulge is spherical and the disk has an exponential vertical distribution with scale height $z_{\rm d} = 0.196\, R_{\rm d}^{0.633}$ \citep{Bershady2010b}. The uncertainties in $V_{\rm bul}$ and $V_{\rm disk}$ are dominated by $\Upsilon_{\star}$ rather than geometry. For example, for oblate bulges the peak in $V_{\rm bul}$ would increase by less than 20$\%$ \citep{Noordermeer2008}. For galaxies with $T\geq4$, we assume that all stars lie in a disk and interpret possible light concentrations as disky pseudobulges \citep{Kormendy2004}. For only five Sbc-to-Scd galaxies (IC~4202, NGC~5005, NGC~5033, NGC~6946, UGC~2885), bulge-disk decompositions improve the results. Conversely, we do not decompose NGC 3769 despite being an Sb.

In Table~\ref{tab:massmodel}, we provide the values of $V_{\rm bul}$ and $V_{\rm disk}$ for $\Upsilon_{\star}=1$ $M_{\odot}/L_{\odot}$. This is just a convenient value but often exceeds a maximum disk fit. $V_{\rm bul}$ and $V_{\rm star}$ can be trivially rescaled to any arbitrary $\Upsilon_{\star}$.

\begin{figure*}[thb]
\centering
\includegraphics[width=0.9\textwidth]{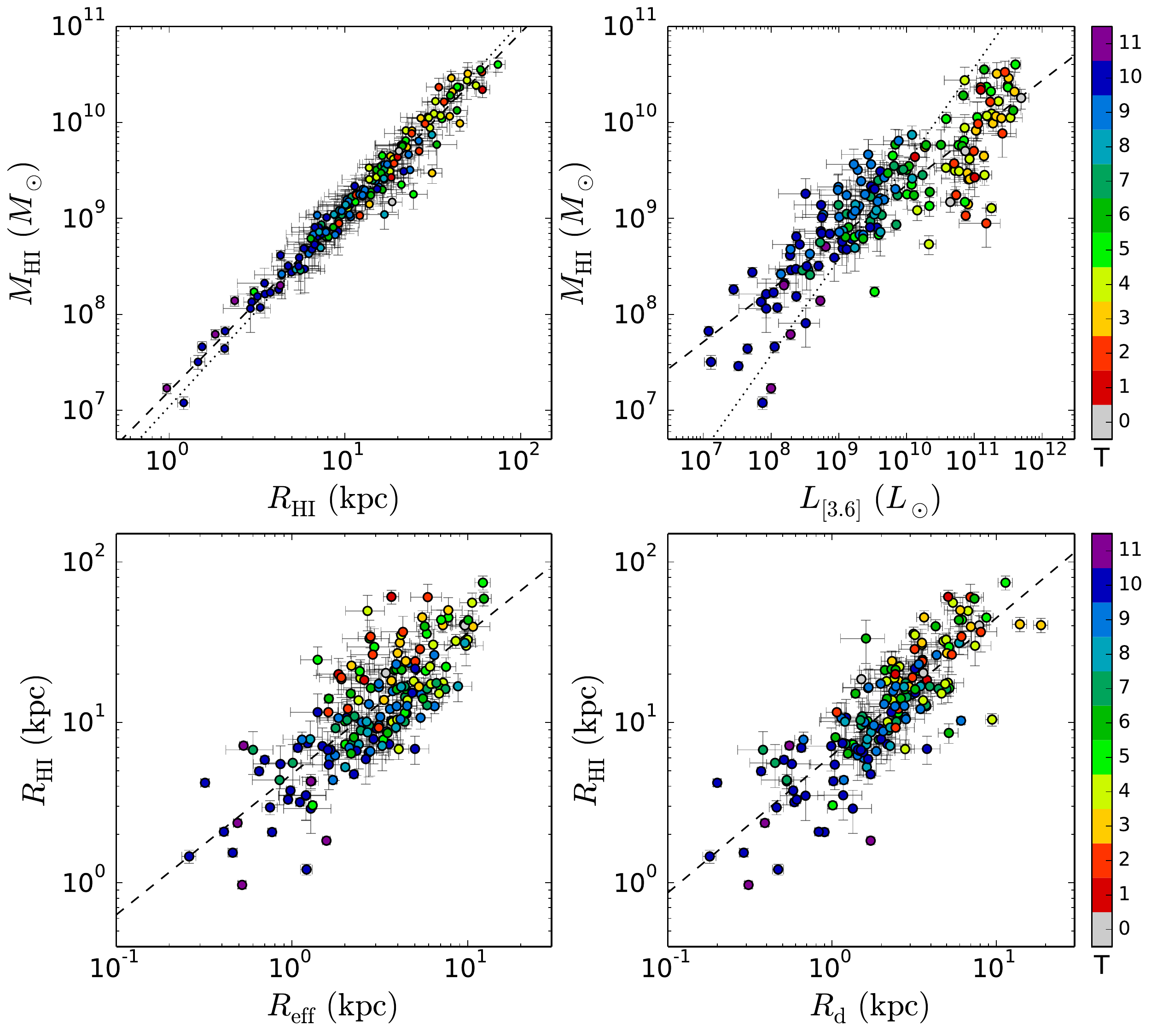}
\caption{Stellar-\hi scaling relations. Galaxies are color-coded by Hubble type. In each panel, the dashed line shows a linear fit. \textit{Top left}: $M_{\hi}$ versus $R_{\hi}$. The dotted line shows the expected relation for \hi disks with a constant mean surface density of 3.5 $M_{\odot}$~pc$^{-2}$. \textit{Top right}: $M_{\hi}$ versus $L_{[3.6]}$. The dotted line shows $M_{\star}=M_{\rm gas}$ for $\Upsilon_{\star}=0.5$ $M_{\odot}/L_{\odot}$. \textit{Bottom}: $R_{\hi}$ versus $R_{\rm eff}$ (\textit{left}) and $R_{\rm d}$ (\textit{right}).}
\label{fig:HIrel}
\end{figure*}
\section{Structural Properties}\label{sec:scaling}

We study structural scaling relations of SPARC galaxies, considering both photometric properties ($L_{[3.6]}$, $R_{\rm eff}$, $\Sigma_{\rm eff}$, $R_{\rm d}$, and $\Sigma_{\rm d}$) and \hi properties ($M_{\hi}$ and $R_{\hi}$).

\subsection{Photometric Scaling Relations}

Figure~\ref{fig:photo} illustrates the covariance between $L_{[3.6]}$, $R_{\rm eff}$, and $\Sigma_{\rm eff}$ (top panels) as well as $L_{[3.6]}$, $R_{\rm d}$, and $\Sigma_{\rm d}$ (bottom panels). As expected, the total luminosity broadly correlates with both radius ($R_{\rm eff}$ and $R_{\rm d}$) and surface brightness ($\Sigma_{\rm eff}$ and $\Sigma_{\rm d}$). In the top-left panel, the solid line shows the $K$-band mass-radius relation from the GAMA survey \citep{Lange2015} for morphologically identified late-type galaxies. For comparison, the stellar masses from \citet{Lange2015} are converted into [3.6] luminosities adopting $\Upsilon_{\star}=0.5$ $M_{\odot}/L_{\odot}$ \citep{Schombert2014a}. The dashed lines show a 0.4 dex deviation from the GAMA relation, corresponding to $\sim$2 times the observed rms scatter ($\sigma_{\rm obs}$). We find overall agreement with \citet{Lange2015}, indicating that SPARC is not missing any type of galaxy disk in terms of luminosity, size, or surface brightness.

In Figure~\ref{fig:photo}, one may discern two separate sequences in the luminosity-size planes, corresponding to early-type HSB spirals and late-type LSB galaxies. This separation may be either real or due to the under-representation of galaxies with $L_{[3.6]}\simeq10^{10}$ $L_{\odot}$ in SPARC. Interestingly, a similar separation has been found by \citet{Schombert2006} and may indicate the existence of a surface brightness bimodality for stellar disks \citep{Tully1997, McDonald2009a, McDonald2009b, Sorce2013b}.

\subsection{The \hi Mass-Radius Relation}

Figure~\ref{fig:HIrel} (top left) shows the $M_{\hi}-R_{\hi}$ relation \citep{Broeils1997, Verheijen2001b}. This correlation is extremely tight. We fit a line using LTS\_LINEFIT \citep{Cappellari2013}, which considers errors in both variables and allows for intrinsic scatter ($\sigma_{\rm int}$). LTS\_LINEFIT assumes that the errors on the two variables are independent, but $M_{\hi}$ and $R_{\hi}$ are affected by galaxy distance as $D^{2}$ and $D$, respectively. Hence, we introduce distance uncertainties only in $M_{\hi}$ with a linear dependence. We also consider typical errors of 10$\%$ on $M_{\hi}$ and $R_{\hi}$ due to flux calibration. We find
\begin{equation}
\log(M_{\hi}) = (1.87 \pm 0.03) \log(R_{\hi}) - (7.20 \pm 0.03),
\end{equation}
with $\sigma_{\rm obs} = 0.13$ dex and $\sigma_{\rm int}=0.06 \pm 0.01$ dex. This is one of the tightest scaling relations of galaxy disks. For comparison, the BTFR has $\sigma_{\rm obs}\simeq0.18$ dex and $\sigma_{\rm int}\lesssim$0.11 dex \citep{Lelli2016}.

A tight $M_{\hi}-R_{\hi}$ relation with a slope of $\sim$2 implies that the mean surface density of \hi disks is constant from galaxy to galaxy \citep{Broeils1997}. For comparison, the dotted line in Figure~\ref{fig:HIrel} shows the expected relation for \hi disks with a mean surface density of 3.5 $M_{\odot}$ pc$^{-2}$. We find a slope of 1.87 which is similar to previous estimates: 1.96 \citep{Broeils1997}, 1.88 \citep{Verheijen2001b}, 1.86 \citep{Swaters2002a}, 1.80 \citep{Noordermeer2005}, and 1.72 \citep{Martinsson2016}. These small differences likely arise from different sample sizes and ranges in galaxy properties: in this respect SPARC is the largest sample and spans the broadest dynamic range. The small tilt from the ``expected'' value of 2 implies that the \hi disks of low-mass galaxies are denser than those of high-mass ones (by a factor of $\sim$2, see also \citealt{Noordermeer2005, Martinsson2016}).

\subsection{Stellar-\hi Scaling Relations}

Figure~\ref{fig:HIrel} (top right) shows the relation between $M_{\hi}$ and $L_{[3.6]}$. A linear fit returns
\begin{equation}
 \log(M_{\hi}) = (0.54 \pm 0.02) \log(L_{[3.6]}) + (3.90 \pm 0.23),
\end{equation}
with $\sigma_{\rm int} = 0.35\pm0.02$ dex. A slope of $\sim$0.6 implies that the ratio of gas mass ($M_{\rm gas} = 1.33 M_{\hi}$) to stellar mass ($M_{\star} = \Upsilon_{\star} L_{[3.6]}$) is not constant in galaxies, but become systematically higher for low-luminosity galaxies. For comparison, the dotted line shows the equality $M_{\rm gas} = M_{\star}$ assuming $\Upsilon_{\star} = 0.5$ $M_{\odot}/L_{\odot}$. Interestingly, this line broadly demarcates the transition from spirals (T = 1 to 7) to irregulars (T = 8 to 11). Gas fractions are explored in detail in Sect.~\ref{sec:gasfrac}.

Finally, the bottom panels of Fig.~\ref{fig:HIrel} show the relation between \hi radius and stellar radius. The latter is estimated as $R_{\rm eff}$ (left) and $R_{\rm d}$ (right). Assuming a 10$\%$ error on these radii, a linear fit returns
\begin{equation}
\log(R_{\hi}) = (0.88 \pm 0.05) \log(R_{\rm eff}) + (0.68 \pm 0.03)
\end{equation}
with $\sigma_{\rm int} = 0.22 \pm 0.01$ dex, and
\begin{equation}
\log(R_{\hi}) = (0.86 \pm 0.04) \log(R_{\rm d}) + (0.79 \pm 0.02), 
\end{equation}
with $\sigma_{\rm int} = 0.20 \pm 0.01$ dex. The slope and normalization of these relations imply that \hi disks are generally more extended than stellar disks, as we discussed in Sect.~\ref{sec:prop}.

\section{Dynamical Properties}\label{sec:results}

In this section, we explore dynamical properties of SPARC galaxies like gas fractions (Sect.~\ref{sec:gasfrac}) and degree of baryonic maximality (Sect.~\ref{sec:maximality}). Both quantities depend on the assumed stellar mass-to-light ratio, hence we start by discussing the plausible values of $\Upsilon_{\star}$ at [3.6].

\subsection{The Stellar Mass-to-Light Ratio}\label{sec:ML}

The major uncertainty in the mass modelling of galaxies is the value of $\Upsilon_{\star}$. To address this issue, the DiskMass Survey \citep[DMS,][]{Bershady2010a} measured the vertical velocity dispersion of disk stars in 30 face-on spirals and employed the theoretical relation between vertical velocity dispersion, disk scale height, and dynamical surface density (valid for a self-graviting disk). Since $z_{\rm d}$ cannot be directly measured in face-on galaxies, the DMS statistically estimated the disk scale height using the empirical correlation between $R_{\rm d}$ and $z_{\rm d}$, calibrated using NIR photometry of edge-on galaxies \citep{Kregel2002}.

The DMS concluded that stellar disks are strongly submaximal, contributing less than 30$\%$ of the dynamical mass within 2.2 $R_{\rm d}$ and having $\Upsilon_{\star}=0.31$ $M_{\odot}/L_{\odot}$ in the $K$-band \citep{Martinsson2013}. After correcting for the DM contribution, the DMS gives $\Upsilon_{\star}=0.24$ $M_{\odot}/L_{\odot}$ in the $K$-band \citep{Swaters2014}. Considering that $L_{K} \simeq 1.3 L_{[3.6]}$ \citep{Schombert2014a}, these values correspond to $\Upsilon_{\star}\simeq0.2$ $M_{\odot}/L_{\odot}$ at [3.6]. \citet{Angus2016} re-analized the DMS data pointing out that the DM contribution to the vertical force become significant for such submaximal disks: when DM haloes are included in a self-consistent way, the mean value of $\Upsilon_{\star}$ further decreases to $\sim$0.18~$M_{\odot}/L_{\odot}$ in the $K$-band ($\sim$0.14 $M_{\odot}/L_{\odot}$ at [3.6]). These results are in tension with SPS models \citep{McGaugh2014, Meidt2014, Schombert2014a}, which find mean values between $\sim$0.6 and $\sim$0.8 $M_{\odot}/L_{\odot}$ in the $K$-band ($\sim$0.4 to $\sim$0.6 $M_{\odot}/L_{\odot}$ at [3.6]) depending on the model and IMF.

Interestingly, both the DMS and SPS models suggest that $\Upsilon_{\star}$ is nearly constant in the NIR (within $\sim$0.1 dex) among galaxies of different masses and morphologies. The discrepancy is only in the overall normalization of $\Upsilon_{\star}$. A nearly constant $\Upsilon_{\star}$ at [3.6] is also suggested by the BTFR \citep{McGaugh2015}, which is remarkably tight for a fixed $\Upsilon_{\star}$ \citep{Lelli2016}. Thus, we assume that $\Upsilon_{\star}$ is constant among different galaxies and explore different normalizations. For 32 galaxies with significant bulges, however, we adopt
\begin{equation}
\Upsilon_{\rm bul} = 1.4 \Upsilon_{\rm disk},
\end{equation}
as suggested by SPS models \citep{Schombert2014a}. Hereafter, the different normalizations of $\Upsilon_{\star}$ always refer to the stellar disk, i.e. $\Upsilon_{\star}=\Upsilon_{\rm disk}$.

\begin{figure*}[thb]
\centering
\includegraphics[width=0.95\textwidth]{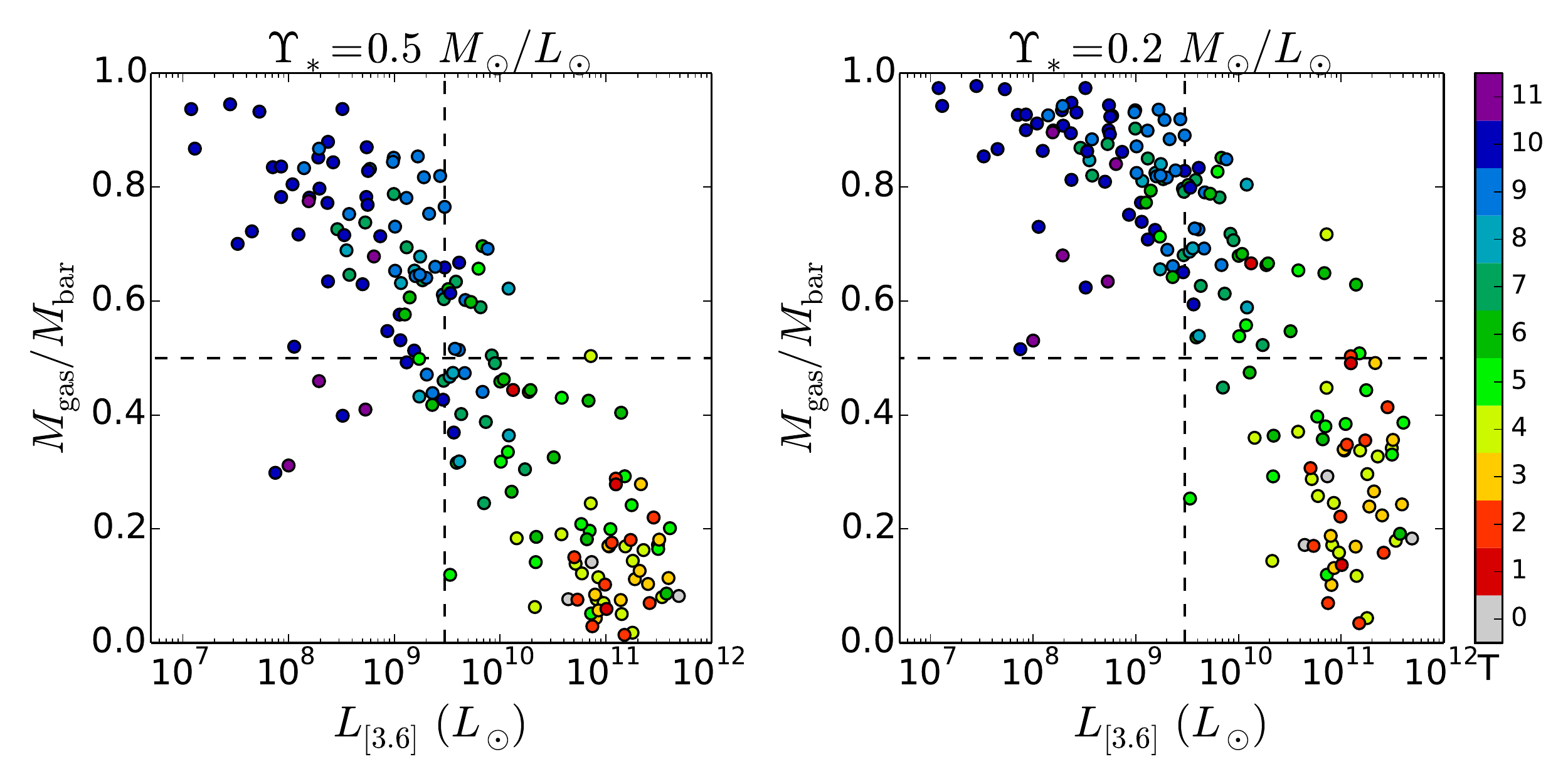}
\caption{Gas fraction (Eq.\,\ref{Eq:fg}) versus [3.6] luminosity for $\Upsilon_{\star} = 0.5$ $M_{\odot}/L_{\odot}$ (\textit{left}) and $\Upsilon_{\star} = 0.2$ $M_{\odot}/L_{\odot}$ (\textit{right}). Galaxies are color-coded by Hubble type. The vertical line at $L_{[3.6]} = 3 \times 10^{9} L_{\odot}$ approximately separates spirals (T = 1 to 7) from dwarf irregulars (T = 8 to 11).}
\label{fig:gasfrac}
\end{figure*}
\subsection{Gas Fractions}\label{sec:gasfrac}

We define the gas fraction as
\begin{equation}\label{Eq:fg}
f_{\rm gas} = \dfrac{M_{\rm gas}}{M_{\rm bar}} = \dfrac{M_{\rm gas}}{M_{\rm gas} + \Upsilon_{\star}L_{\rm disk} + 1.4\Upsilon_{\star} L_{\rm bul}},
\end{equation}
where $L_{\rm disk}$ and $L_{\rm bul}$ are estimated using the non-parametric decompositions in Sect.\,\ref{sec:massmodels}.
Figure~\ref{fig:gasfrac} plots $f_{\rm gas}$ versus $L_{[3.6]}$ assuming $\Upsilon_{\star} = 0.5$~$M_{\odot}/L_{\odot}$ (left) and $\Upsilon_{\star} = 0.2$~$M_{\odot}/L_{\odot}$ (right). For any realistic value of $\Upsilon_{\star}$, the gas fraction anticorrelates with $L_{[3.6]}$ but the overall shapes of these relations significantly change with $\Upsilon_{\star}$.

A value of $\Upsilon_{\star}\simeq0.5$~$M_{\odot}/L_{\odot}$ minimizes the scatter around the BTFR \citep{Lelli2016} and is found using self-consistent SPS models with standard IMFs \citep{McGaugh2014, Schombert2014a}. This normalization leads to the usual picture for the composition of galaxy disks:
\begin{enumerate}
 \item The $f_{\rm gas}-\log(L_{[3.6]})$ relation is roughly linear. Dwarf galaxies are generally gas-dominated but show a broad range in $f_{\rm gas}$ from $\sim$0.5 to $\sim$0.9.
 \item The transition between spirals (T = 1 to 7) and irregulars (T = 8 to 11) roughly occurs at $f_{\rm gas}\simeq0.5$. This is in line with density wave theory \citep{Lin1964}, since well-developed spiral patterns are difficult to support in gas-dominated disks.
\end{enumerate}

A value of $\Upsilon_{\star}=0.2$~$M_{\odot}/L_{\odot}$ at [3.6] is comparable to the DMS normalization. This low value of $\Upsilon_{\star}$ leads to some strange results:
\begin{enumerate}
 \item The $f_{\rm gas}-\log(L_{[3.6]})$ relation shows a flattening at $L_{[3.6]} \lesssim 3 \times 10^{9} L_{\odot}$ (vertical dashed line). Consequently, the vast majority of dwarf irregulars are heavily gas-dominated with $f_{\rm gas} \simeq 0.8-1.0$. Dwarf irregulars are generally thought to be gas-dominated \citep[e.g.,][]{McGaugh2012}, but this seems a very extreme situation.
 \item Several bright spirals ($10^{10} < L_{[3.6]}/L_{\odot} < 10^{11}$) have $f_{\rm gas} \gtrsim 0.5$. This also includes four early-type spirals (Sa to Sb), corresponding to $T = 1$ to 3. It is unclear how these bright galaxies could sustain stellar bars and grand-design spiral patterns in such gas-dominated disks \citep[see e.g.,][]{Athanassoula1984}.
\end{enumerate}

This suggests that $\Upsilon_{\star}\simeq0.5$ $M_{\odot}/L_{\odot}$ gives a more realistic galaxy population than $\Upsilon_{\star}\simeq0.2$ $M_{\odot}/L_{\odot}$.

\subsection{Baryonic Maximality}\label{sec:maximality}

Historically, the degree of disk maximality is estimated by measuring the ratio $V_{\star}/V_{\rm obs}$ at 2.2$R_{\rm d}$ \citep{Sackett1997}. This choice is inherited from early works on late-type spiral galaxies \citep[e.g.,][]{vanAlbada1985} and is motivated by two facts: (i) in spiral galaxies $V_{\rm gas}$ is typically much smaller than $V_{\star}$, and (ii) for a pure exponential disk $V_{\star}$ peaks at 2.2$R_{\rm d}$. When comparing galaxies of different masses and morphologies, however, this choice becomes inadequate because (i) in dwarf galaxies $V_{\rm gas}$ can be larger than $V_{\star}$, and (ii) the vast majority of spiral galaxies show deviations from exponential profiles in the inner parts due to bars, lenses, and bulges. Hence, we consider the ratio $V_{\rm bar}/V_{\rm obs}$ (instead of $V_{\star}/V_{\rm obs}$) and explore several characteristic radii.

\begin{figure*}[t!]
\begin{minipage}{0.5\textwidth}
\centering
\includegraphics[width=\textwidth]{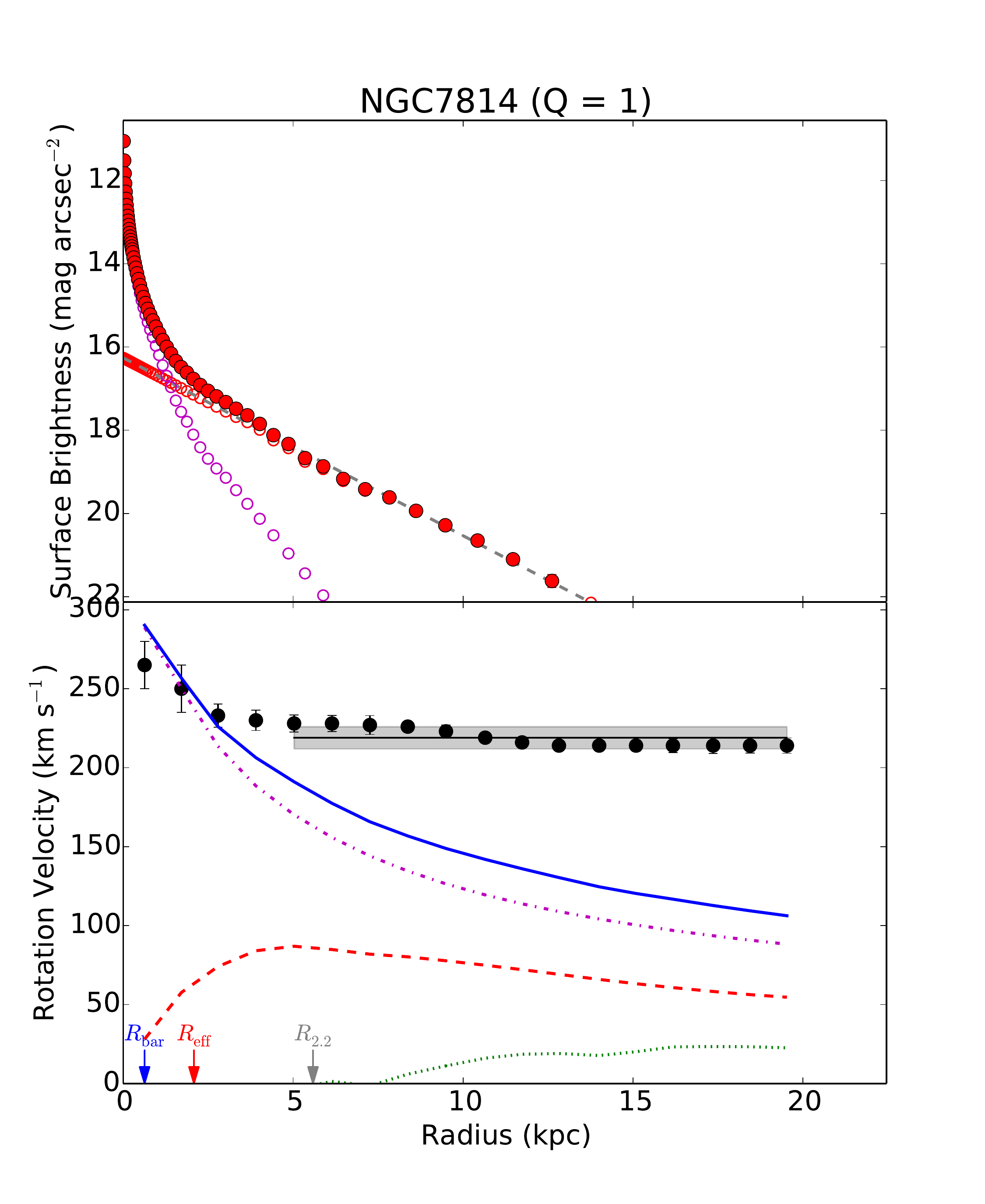}
\end{minipage}
\begin{minipage}{0.5\textwidth}
\centering
\includegraphics[width=\textwidth]{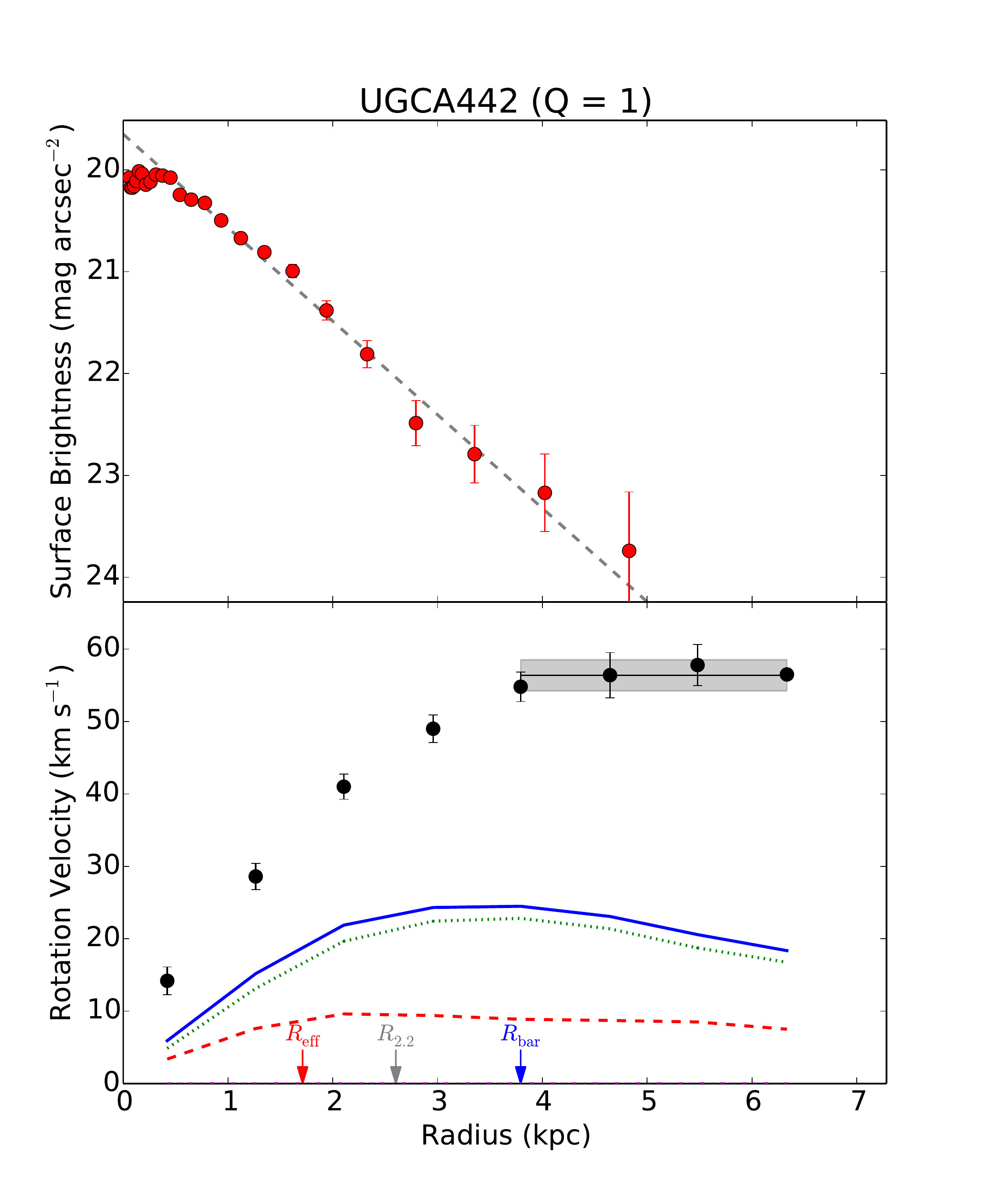}
\end{minipage}
\caption{Mass models for two SPARC galaxies adopting $\Upsilon_{\rm disk} = 0.5$ $M_{\odot}$/$L_{\odot}$ and $\Upsilon_{\rm bul} = 0.7$ $M_{\odot}$/$L_{\odot}$. Top panels show the [3.6] surface brightness profiles (red dots) and exponential fits to the outer stellar disk (dashed line). Open circles show extrapolated values for bulge (purple) and disk (red) components. Bottom panels show the observed rotation curves (black dots) and the velocity contributions due to gas (green dotted line), stars (red dashed line), bulge (purple dash-dotted line), and total baryons (blue solid line). The arrows indicate three characteristic radii: 2.2 $R_{\rm d}$ (grey), $R_{\rm eff}$ (red), and $R_{\rm bar}$ (blue). The grey band shows the velocity-points used to compute $V_{\rm f}$ (black-solid line), as described in \citet{Lelli2016}. Similar figures for all SPARC galaxies are available as Figure Set.}
\label{fig:massmodels}
\end{figure*}
\subsubsection{The Choice of the Characteristic Radius}

Figure~\ref{fig:massmodels} shows mass models for two galaxies with diametrically opposed properties: a bulge-dominated spiral galaxy (NGC\,7814) and a gas-dominated dwarf galaxy (UGCA\,442). Top panels show [3.6] surface brightness profiles and exponential fits to the outer disk; bottom panels show rotation curve decompositions adopting $\Upsilon_{\rm disk}=0.5$ $M_{\odot}$/$L_{\odot}$ and $\Upsilon_{\rm bul}=0.7$ $M_{\odot}$/$L_{\odot}$. The arrows illustrate three different characteristic radii: (i)~the commonly used radius at 2.2$R_{\rm d}$ ($R_{2.2}$, grey); (ii)~the effective radius encompassing half of the total [3.6] luminosity ($R_{\rm eff}$, red); and (iii) the radius where $V_{\rm bar}$ is maximum ($R_{\rm bar}$, blue). The last radius was denoted as $R_{\rm p}$ by \citet{McGaugh2005b}.

NGC~7814 (left) is an early-type spiral (Sab). The surface brightness profile shows a central light enanchement due to the bulge. The rotation curve displays a very steep rise in the inner parts, a decline of $\sim$50 km~s$^{-1}$, and a flat outer portion. These are typical features for early-type spirals \citep{Noordermeer2007}. For $\Upsilon_{\star}=0.5$ $M_{\odot}$/$L_{\odot}$ baryons explain the inner parts of the rotation curve, while DM is needed at large radii. Note that $R_{\rm eff}$ occurs along the declining part of the rotation curve, while $R_{2.2}$ occurs along the flat part: $R_{\rm bar}$ is the only radius that captures the concept of baryonic maximality.

UGCA~442 (right) is a late-type dwarf (Sm). The surface brightness profile is well described by an exponential law, but there is a central flattening due to a cored light distribution. The rotation curve shows a slow rise in the central regions and a flat part beyond the stellar component. For $\Upsilon_{\star}=0.5$ $M_{\odot}$/$L_{\odot}$ baryons cannot fully explain the inner rise of the rotation curve and DM is needed down to small radii. This is a typical behaviour for low-mass and LSB galaxies: maximum disk decompositions require high values of $\Upsilon_{\star}$ that are ruled out by standard SPS models \citep{deBlok1997a, Swaters2011}. Note that the shape of $V_{\rm bar}$ is mostly driven by the gas contribution and $R_{\rm bar}$ occurs well beyond $R_{\rm eff}$ or $R_{2.2}$ due to the gas dominance.

These two examples illustrate that both $R_{2.2}$ and $R_{\rm eff}$ can be misleading in quantifying the baryonic maximality because they can be far from the radius where $V_{\rm bar}$ peaks. In the following, we consider $R_{\rm bar}$ as our fiducial characteristic radius but we also show $V_{\rm bar}/V_{\rm obs}$ at $R_{2.2}$ (hereafter $V_{2.2}/V_{\rm obs}$) to compare with previous works.

In the previous examples, the value of $R_{\rm bar}$ does not depend on the assumed $\Upsilon_{\star}$ because the shape of $V_{\rm bar}$ is dominated by either $V_{\star}$ or $V_{\rm gas}$. In galaxies where $V_{\star}\simeq V_{\rm gas}$, however, the value of $R_{\rm bar}$ may depend on the assumed $\Upsilon_{\star}$ because the overall shape of $V_{\rm bar}$ is affected by the relative contributions of $V_{\star}$ and $V_{\rm gas}$: these cases represent a minority among SPARC galaxies.

Finally, we note that $V_{\rm bar}$ depends on galaxy distance as $\sqrt{D}$ other than on the assumed $\Upsilon_{\star}$. Uncertainties in galaxy distances, therefore, may affect the value of $V_{\rm bar}/V_{\rm obs}$. Moreover, the normalization of $V_{\rm obs}$ depends on the assumed inclination as $\sin(i)$, but this is a minor source of uncertainty for galaxies with $i\gtrsim30^{\circ}$.

\begin{figure*}[thb]
\centering
\includegraphics[width=0.9\textwidth]{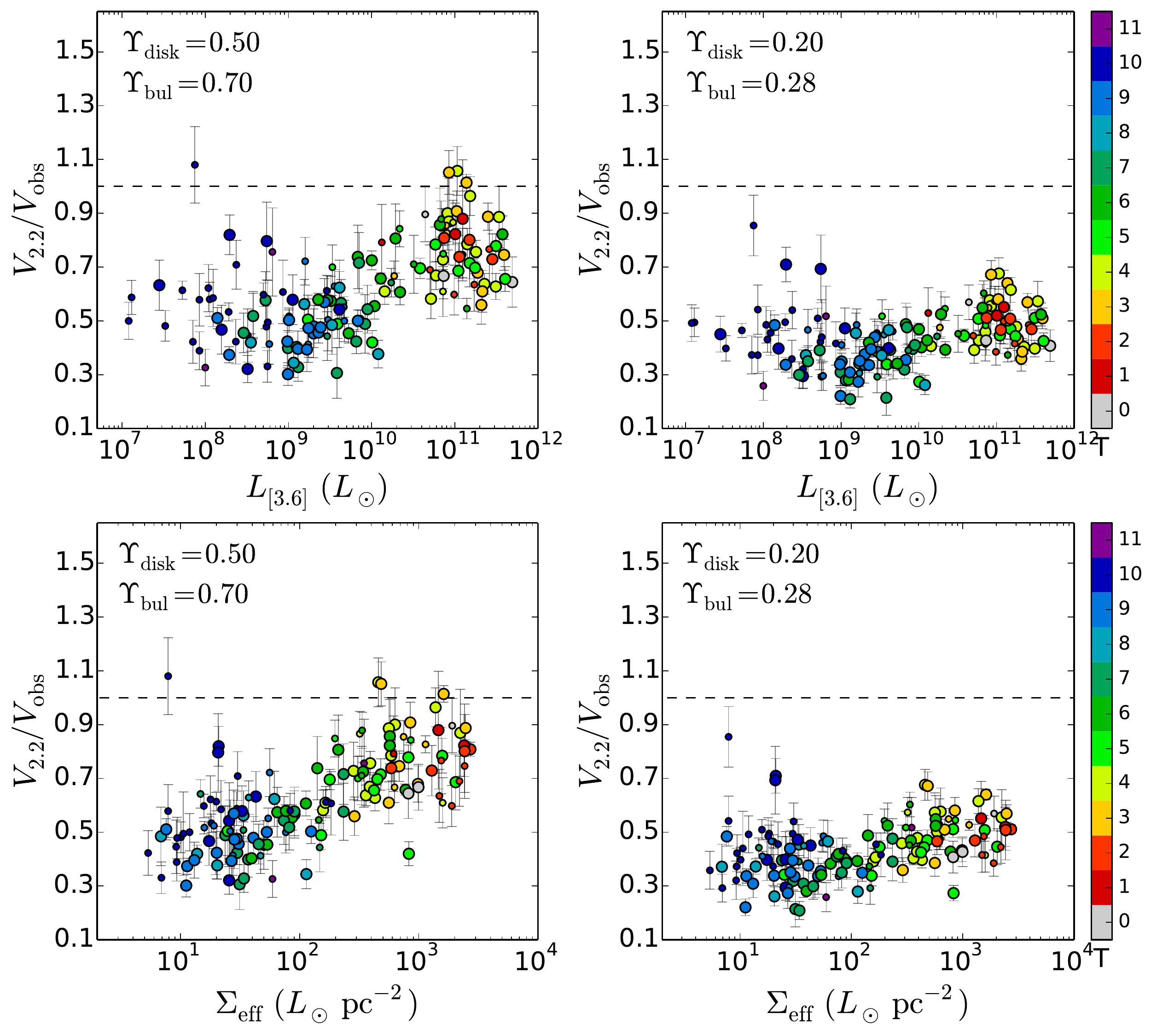}
\caption{Ratio of baryonic to observed rotation velocity at 2.2 disk scale lengths ($V_{2.2}/V_{\rm obs}$). We show the covariance with the total [3.6] luminosity (top) and effective surface brightness (bottom). Two different normalizations of $\Upsilon_{\star}$ are illustrated. Galaxies are color-coded by Hubble type. In these plots we exclude galaxies with $i<30^{\circ}$ and $Q=3$. Large and small circles correspond to galaxies with $Q=1$ and $Q=2$, respectively. The dashed line shows $V_{\rm bar}/V_{\rm obs} = 1$, denoting the upper limit for physically acceptable disks.}
\label{fig:ML1}
\end{figure*}
\begin{figure*}[thb]
\centering
\includegraphics[width=0.9\textwidth]{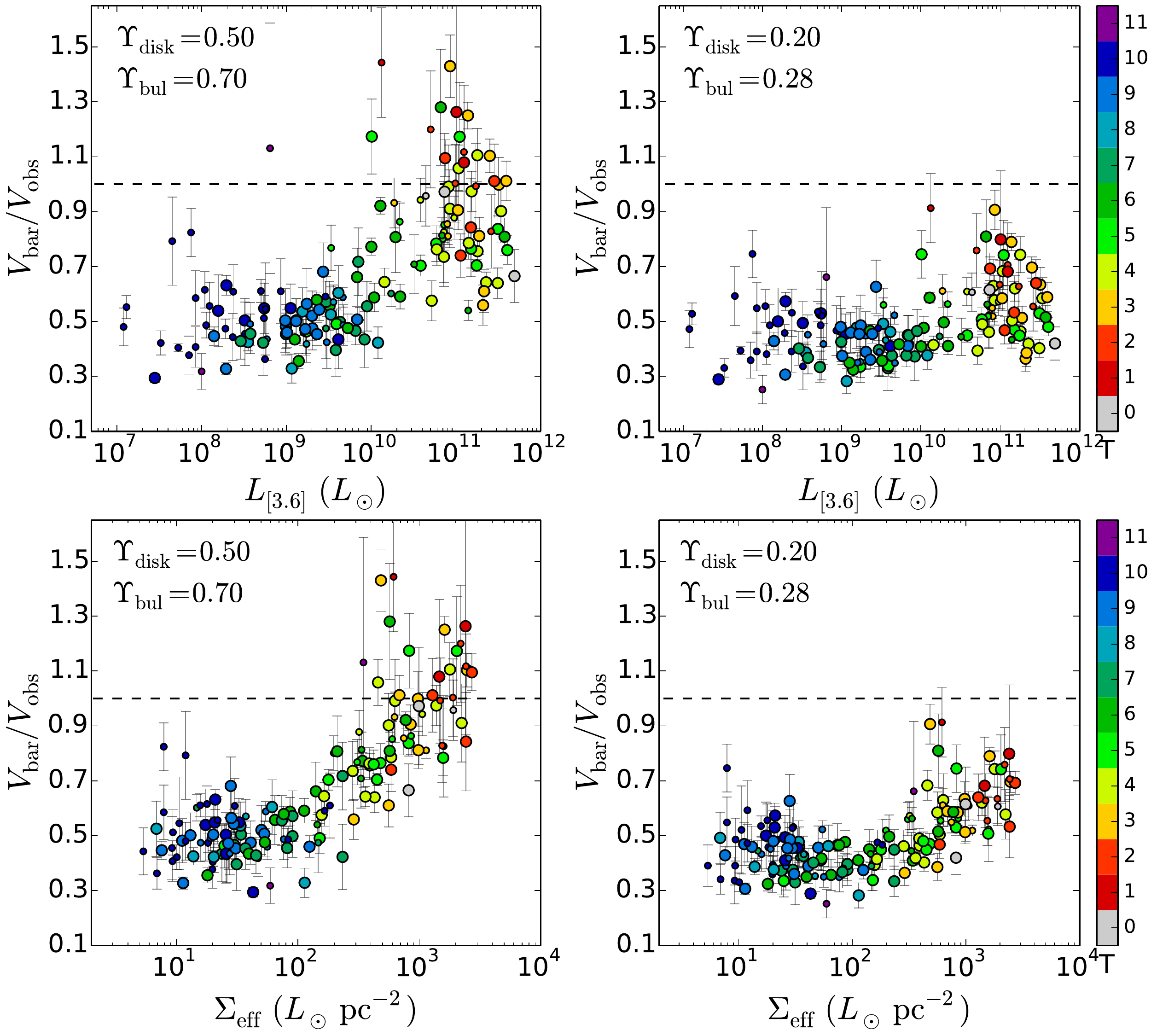}
\caption{Same as Figure\,\ref{fig:ML1} but we now show the value of $V_{\rm bar}/V_{\rm obs}$ measured at the peak of the baryonic contribution ($R_{\rm bar}$).}
\label{fig:ML2}
\end{figure*}
\begin{figure}[thb]
\centering
\includegraphics[width=0.47\textwidth]{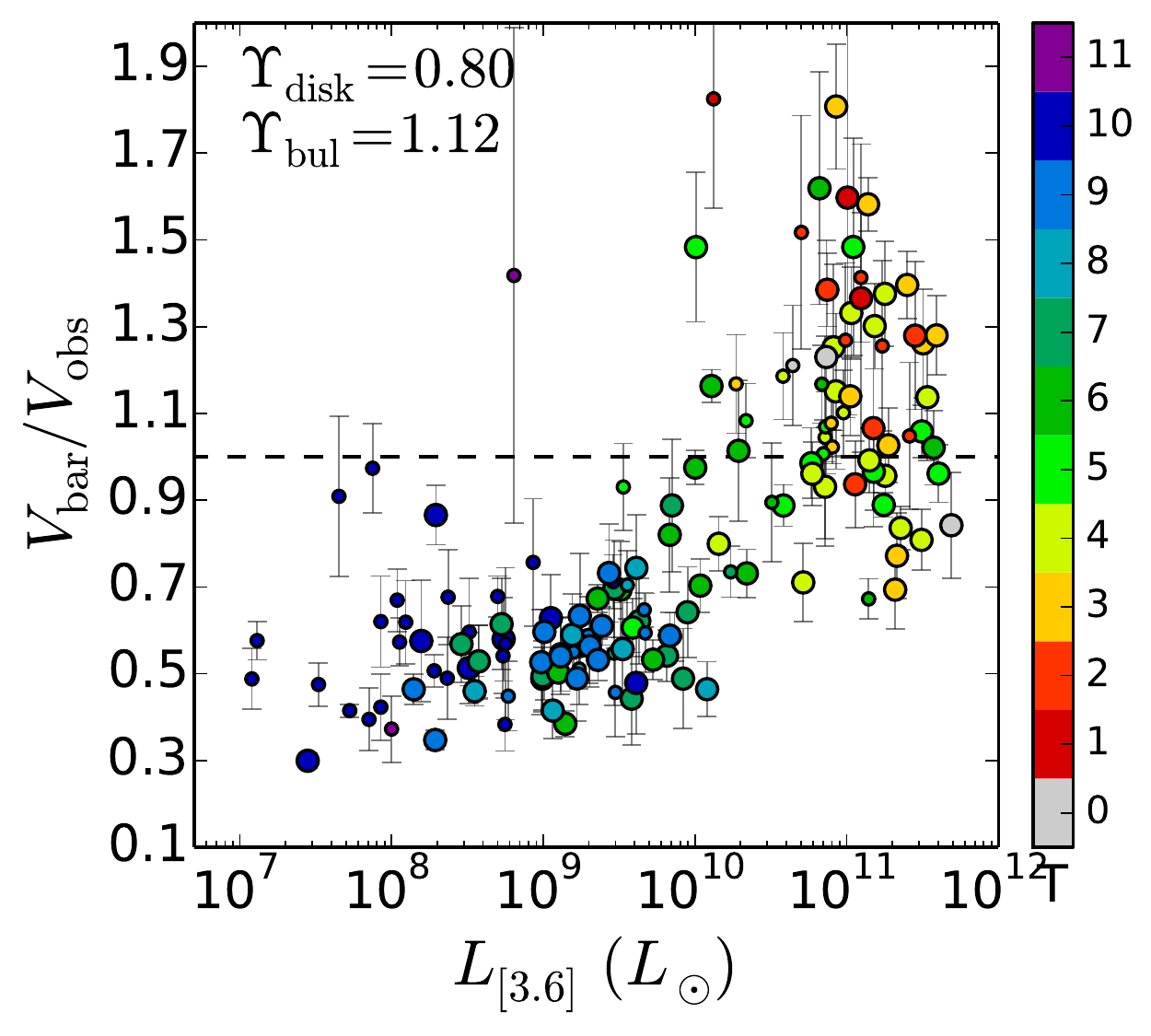}
\caption{Same as Figure~\ref{fig:ML2} (top) but adopting $\Upsilon_{\star}=0.8$~$M_{\odot}/L_{\odot}$. This value of $\Upsilon_{\star}$ gives unphysical results for high-mass galaxies.}
\label{fig:ML3}
\end{figure}
\subsubsection{The Effect of the Stellar Mass-To-Light Ratio}

Figures \ref{fig:ML1} and \ref{fig:ML2} show $V_{\rm bar}/V_{\rm obs}$ measured at $R_{2.2}$ and $R_{\rm bar}$, respectively. We illustrate the covariance with $L_{[3.6]}$ (top) and $\Sigma_{\rm eff}$ (bottom). We consider two different normalizations of $\Upsilon_{\star}$ as described in Sect.\,\ref{sec:ML}. Firstly, we note that the degree of maximality of dwarf galaxies does not strongly depend on $\Upsilon_{\star}$ because these objects are gas-dominated: the values of $V_{\rm bar}/V_{\rm obs}$ stay in the range 0.3 to 0.6 (apart a few outliers) using both $R_{2.2}$ and $R_{\rm bar}$. Conversely, the degree of maximality of spirals depends strongly on $\Upsilon_{\star}$.

For $\Upsilon_{\star} = 0.5$ $M_{\odot}/L_{\odot}$, high-luminosity HSB galaxies are nearly maximal, whereas low-luminosity LSB galaxies are submaximal. By definition, the value of $V_{\rm bar}/V_{\rm obs}$ is higher at $R_{\rm bar}$ than $R_{2.2}$, especially in galaxies with bulges. Some objects have $V_{\rm bar}/V_{\rm obs} \gtrsim 1$ but this is not surprising. We expect some intrinsic scatter around the mean value of $\Upsilon_{\star}$ \citep[of $\sim$0.1 dex, see][]{Meidt2014, Lelli2016}, hence a few objects may have $V_{\rm bar}/V_{\rm obs}>1$ for a fixed $\Upsilon_{\star}$. We also note that several galaxies with $V_{\rm bar}/V_{\rm obs}>1$ have uncertain distances and are consistent with $V_{\rm bar}/V_{\rm obs} \lesssim 1$ within the errors. The degree of baryonic maximality correlates better with surface brightness than luminosity, in line with previous studies \citep[e.g.,][]{McGaugh2005b, McGaugh2016}.

For $\Upsilon_{\star} = 0.2$ $M_{\odot}/L_{\odot}$, spiral galaxies are submaximal as found by the DMS \citep{Martinsson2013}. For this low value of $\Upsilon_{\star}$, most spirals are nearly as submaximal as dwarf galaxies (which are not included in the DMS). This would imply that there is no strong dynamical difference among objects spanning $\sim$5 dex in luminosity and $\sim$4 dex in surface brightness. One does expect a difference over this range because HSB and LSB galaxies lie on the same BTFR. At a given baryonic mass, $V_{\rm obs}$ is nearly the same but $V_{\rm bar}$ goes as $\sqrt{M_{\rm bar}/R_{\rm bar}}$, hence it is higher in HSB galaxies (with small $R_{\rm bar}$) than in LSB galaxies (with large $R_{\rm bar}$). Therefore, $V_{\rm bar}/V_{\rm obs}$ should correlate with surface brightness, but this trend is barely perceptible for $\Upsilon_{\star}=0.2$ $M_{\odot}/L_{\odot}$. If $R_{\rm bar}$ is adopted, $V_{\rm bar}/V_{\rm obs}$ shows a parabolic trend with $\Sigma_{\rm eff}$ (Figure\,\ref{fig:ML2}, bottom right). The LSB end is driven by gas-dominated dwarf galaxies, while the HSB end is driven by bulge-dominated spiral galaxies: galaxies with intermediate surface brightnesses are generally disk-dominated and would be less maximal than gas-dominated LSB galaxies for $\Upsilon_{\star}=0.2$ $M_{\odot}/L_{\odot}$. This seems strange.

It is important to derive an upper limit for the mean value of $\Upsilon_{\star}$ since this can be used to constrain different IMFs and SPS models. In Figure~\ref{fig:ML3} we show $V_{\rm bar}/V_{\rm obs}$ at $R_{\rm bar}$ for $\Upsilon_{\star} = 0.8$ $M_{\odot}/L_{\odot}$. For this high normalization, a large fraction of high-mass HSB galaxies have $V_{\rm bar}/V_{\rm obs} > 1$. It is conceivable that $\Upsilon_{\star}$ may scatter up to 0.8 $M_{\odot}/L_{\odot}$ for some individual galaxies, but this cannot be the correct mean $\Upsilon_{\star}$. In the extreme scenario where the ``true'' mean $\Upsilon_{\star}$ gives $V_{\rm bar}/V_{\rm obs} = 1$ in bright galaxies, the scatter in $\Upsilon_{\star}$ should drive $\sim$50$\%$ of galaxies to values of $V_{\rm bar}/V_{\rm obs} > 1$. If we consider galaxies with $L_{[3.6]}>2\times10^{10} L_{\odot}$, a value of $\Upsilon_{\star} = 0.8$ $M_{\odot}/L_{\odot}$ implies over-maximal disks in $\sim$67$\%$ of the cases, which is statistically unacceptable. For truly maximal disks with $V_{\rm bar}/V_{\rm obs} = 1$, we find that the upper limit on the mean value of $\Upsilon_{\star}$ is $\sim$0.7 $M_{\odot}/L_{\odot}$. If we adopt the customary definition of maximum disk as $V_{\rm bar}/V_{\rm obs} = 0.85$ \citep{Sackett1997}, then the mean maximum-disk value of $\Upsilon_{\star}$ is $\sim$0.5 $M_{\odot}/L_{\odot}$, which is similar to the predictions of SPS models with standard IMFs \citep[e.g.,][]{McGaugh2014, Meidt2014, Schombert2014a}.

\section{Summary \& Conclusions}

We introduce SPARC: a sample of 175 galaxies with new Spitzer photometry at 3.6 $\mu$m and high-quality rotation curves from previous \hiA/H$\alpha$ works. We describe the galaxy sample, derive homogeneous [3.6] photometry, and study structural relations of stellar and \hi disks. We find that the relation between stellar and \hi mass has significant intrinsic scatter ($\sigma_{\rm int}\simeq0.35$ dex) as well as the relation between stellar and \hi radius ($\sigma_{\rm int}\simeq0.2$ dex). Conversely, the \hi mass-radius relation is extremely tight ($\sigma_{\rm int}\simeq0.06$ dex) and implies that \hi disks have the same mean surface density within a factor of $\sim$2.

We build detailed mass models, which are available at \url{astroweb.cwru.edu/SPARC}. In this first SPARC paper, we investigate gas fractions ($f_{\rm gas}=M_{\rm gas}/M_{\rm bar}$) and the degree of baryonic maximality ($V_{\rm bar}/V_{\rm obs}$) for different values of the stellar mass-to-light ratio ($\Upsilon_{\star}$).

Assuming $\Upsilon_{\star}=0.5$~$M_{\odot}/L_{\odot}$, we find the usual picture for the overall composition of galaxy disks:
\begin{enumerate}
 \item The relation between $f_{\rm gas}$ and $\log(L_{[3.6]})$ is roughly linear: dwarf galaxies are generally gas-dominated but display a broad range in $f_{\rm gas}$ from 0.5 to 0.9.
 \item The transition between spiral galaxies and dwarf irregulars roughly occurs at $f_{\rm gas} \simeq 0.5$ in line with the predictions of density wave theory: stellar bars and grand-desing spiral arms are generally suppressed in gas-dominated disks.
 \item The ratio $V_{\rm bar}/V_{\rm obs}$ (measured at the peak of $V_{\rm bar}$) varies with luminosity and surface brightness: high-mass HSB galaxies are nearly maximal, while low-mass LSB galaxies are submaximal. The degree of baryonic maximality correlates better with surface brightness than luminosity.
\end{enumerate}

A mean value of $\Upsilon_{\star}\gtrsim0.7$~$M_{\odot}/L_{\odot}$ is statistically ruled out because it implies over-maximal baryonic contributions in high-mass HSB galaxies. If we adopt the customary definition of maximum disk as $V_{\rm bar}/V_{\rm obs}=0.85$ \citep{Sackett1997}, the mean maximum-disk value of $\Upsilon_{\star}$ is $\sim$0.5 $M_{\odot}/L_{\odot}$ for bright galaxies.

If the assumed value of $\Upsilon_{\star}$ is too low (like 0.2 $M_{\odot}/L_{\odot}$), we find some strange results:
\begin{enumerate}
 \item The $f_{\rm gas}-\log(L_{[3.6]})$ relation shows a flattening at the low-luminosity end, implying that dwarf irregulars are extremely gas-dominated ($f_{\rm gas} \simeq 0.8-1.0$).
 \item Several spiral galaxies (including Sa-Sb types) would be gas-dominated ($f_{\rm gas} \gtrsim 0.5$). It would then be unclear how these spiral galaxies could sustain stellar bars and grand-design spiral patterns in such gas-dominated disks.
 \item The ratio $V_{\rm bar}/V_{\rm obs}$ would show a parabolic trend with surface brightness. Disk-dominated galaxies of intermediate surface brightnesses would be more submaximal than gas-dominated LSB dwarfs as well as bulge-dominated HSB spirals.
\end{enumerate}

A value of $\Upsilon_{\star}=0.2$ $M_{\odot}/L_{\odot}$ at [3.6] ($\sim$0.26 $M_{\odot}/L_{\odot}$ in the $K$-band) is comparable to the results of the DMS \citep{Martinsson2013, Swaters2014}. Hence, one may wonder whether there are some systematics in the DMS or we should change our general view of the composition of galaxy disks. A possible solution to this controversy has been offered by \citet{Aniyan2016}. They pointed out that the scale-heights from the DMS are representative of old stellar populations, whereas the vertical velocity dispersions come from integrated light measurements containing contributions from K-giants of different ages. In the solar vicinity, \citet{Aniyan2016} find that young K-giants have significantly smaller velocity dispersion and scale-height than old K-giants, hence the DMS may have systematically underestimated the disk surface densities by a factor of $\sim$2 due to the combination of light-integrated velocity dispersions from multiple stellar populations and scale-heights from old stellar populations. If true, this would give $\Upsilon_{\star}\simeq0.4-0.5$~$M_{\odot}/L_{\odot}$ in agreement with stellar population synthesis models \citep{McGaugh2014, Meidt2014}, resolved stellar populations in the LMC \citep{Eskew2012}, the BTFR \citep{McGaugh2015, Lelli2016}, and the results presented in this paper.\newline

This work would have not been possible without the high-quality rotation curves that were derived over the past 30 years by many good people. In particular we thank Erwin de Blok, Filippo Fraternali, Gianfranco Gentile, Rachel Kuzio de Naray, Renzo Sancisi, Bob Sanders, Rob Swaters, Thijs van der Hulst, Liese van Zee, and Marc Verheijen. This publication was made possible through the support of a grant from the John Templeton  Foundation. The opinions expressed in this publication are those of the authors and do not necessarily reflect the views of the John Templeton Foundation.

\bibliography{biblio}

\end{document}